# Metal-insulator phase separation in $KTaO_3$-based two-dimensional electron gas

*Jinlei Zhang[1,#], Jiayong Zhang[1,#], Dapeng Cui[2,#], Li Ye[1,#], Shuainan Gong[1], Zhichao Wang[1], Zhenping Wu[3], Chunlan Ma[1], Ju Gao[1,4], Yuanyuan Zhao[5,]\* and Yucheng Jiang[1,]\**

[1]Jiangsu Key Laboratory of Micro and Nano Heat Fluid Flow Technology and Energy Application, School of Physical Science and Technology, Suzhou University of Science and Technology, Suzhou, 215009, China

[2]Department of Physics, Faculty of Science, National University of Singapore, Singapore 117551, Singapore

[3]State Key Laboratory of Information Photonics and Optical Communications & School of Science, Beijing University of Posts and Telecommunications, Beijing 100876, China

[4]School for Optoelectronic Engineering, Zaozhuang University, Shandong 277160, China

[5] School of Physics and Optoelectronic Engineering, Nanjing University of Information Science and Technology, Nanjing 210044, China.

#These authors contributed equally to this work.

*Email: yyzhao@nuist.edu.cn and jyc@usts.edu.cn

# Abstract

Electronic phase separation (EPS) originates from an incomplete transformation between electronic phases, causing the inhomogeneous spatial distribution of electronic properties. In the system of two-dimensional electron gas (2DEG), the EPS is usually identified based on a percolative metal-to-superconductor transition. Here, we report a metal-insulator transition (MIT) in $KTaO_3$-based 2DEG with the width of conductive channel decreasing into micrometer scale. Hysteretic resistance-temperature relations are observed due to the competition between metallic and insulating phases, which is tunable by magnetic field. Such a size-dependent MIT effect is attributed to the coexistence and separation of metallic and insulating phases. Combining density functional theory calculation, we propose a theoretical model to simulate the dynamic process of the EPS using the percolation theory, demonstrating the mechanism of size-dependent MIT. Our work suggests a clear and simple 2DEG platform to achieve the spatial coexistence of metallic and insulating phases.

# INTRODUCTION

Electronic phase separation (EPS) is a ubiquitous physical effect, where micro- to nano-scale regions can be spontaneously formed with distinct electronic, magnetic or conductive states[1,2]. EPS is widely observed in manganites, high-Tc superconductors, transition metal oxides, and so on[3-6]. The instability of ground states plays an important role in the EPS phenomenon[1]. It can thus lead to the spatial coexistence of multiple competing electronic or magnetic states due to their close thermodynamic energies. Usually, the control of phase-separated states can be realized through chemical doping, dimension manipulation and strain engineering[3,7-9]. In complex oxides, the introductions of disorder and oxygen vacancies are the important paths to induce the EPS and tune the length scale of phases[10-13].

Two-dimensional electron gas (2DEG), formed at oxide interfaces, has drawn extensive interest for its intriguing physical properties. In LaAlO$_3$/SrTiO$_3$ systems, the EPS has been reported in a percolative metal-to-superconductor transition, as well as the coexistence of ferromagnetism and superconductivity[12,14]. Recently, another family of 2DEG systems has come into view based on a perovskite oxide of KTaO$_3$ (KTO), including LaAlO$_3$/KTO, LaTiO$_3$/KTO, EuO/KTO and Ar-bombarded KTO, showing the intriguing effects of superconductivity, Rashba effect, time-dependent resistance, etc[15-20]. Generally, oxygen vacancies play an important role in the EPS effect of 2DEG systems[12]. By Ar$^+$-ion bombardment, one can generate oxygen vacancies and introduce the disorder on a KTO surface, offering the possibility to achieve the metal-insulator transition (MIT). Until now, many mechanisms have been found to induce the MIT phenomena in various systems[21-24]. Among them, the percolation-type model plays an important role in the MIT of 2D systems[25,26].

In this letter, we fabricated the 2DEG channels with different widths on KTO substrates using Ar+-ion bombardment and lithography. An effect of MIT was observed as the width decreases. This size-dependent phenomenon can be understood by the EPS. A competition between metallic and insulating phases was reflected by hysteretic resistance-temperature relations. Magnetic field was found to affect the MIT process. Based on our density functional theory (DFT) calculations, the oxygen vacancies on $KTaO_3$ surface cause finite density of states in the vicinity of the Fermi surface, forming metallic spanning-clusters. Thus, we propose a percolation-type theoretical model to simulate the electronic transport, showing good agreement with our experimental results. Our work demonstrates the important role of percolation in achieving 2DEG MIT in KTO, which can be extended to other complex oxide systems.

We prepared the 2DEG channels with various widths on $KTaO_3$ (KTO) substrates using lithography technique and Ar+-ion bombardment as described in Fig. 1a (see Methods in the Supplemental Material). KTO is a wide-bandgap transparent insulator with perovskite lattice structure. A 2DEG can be created through the Ar+-ion bombardment on the KTO substrate. This method has been used to prepare electron gas on the surface of $SrTiO_3$ crystals[27]. An $Ar^+$-ion beam, driven by voltage, can ablate more oxygen ions than metal ions from the KTO substrate, thus forming oxygen vacancies on the surface. The 2DEG is identified as a conductive nanoscale layer. Atomic force microscopy (AFM) images show the four 2DEG channels with the widths from 1 to 10 μm in Fig. 1b. They were simultaneously prepared on the same KTO substrate, to ensure the consistency of preparation conditions. The lattice structure of KTO was measured by a spherical aberration–corrected scanning transmission electron microscopy (Cs-corrected STEM). Fig. 1c exhibits the perovskite structure of KTO in atomic scale, which is verified by its fast Fourier transform (FFT) pattern. This also implies the high quality of KTO single crystal.

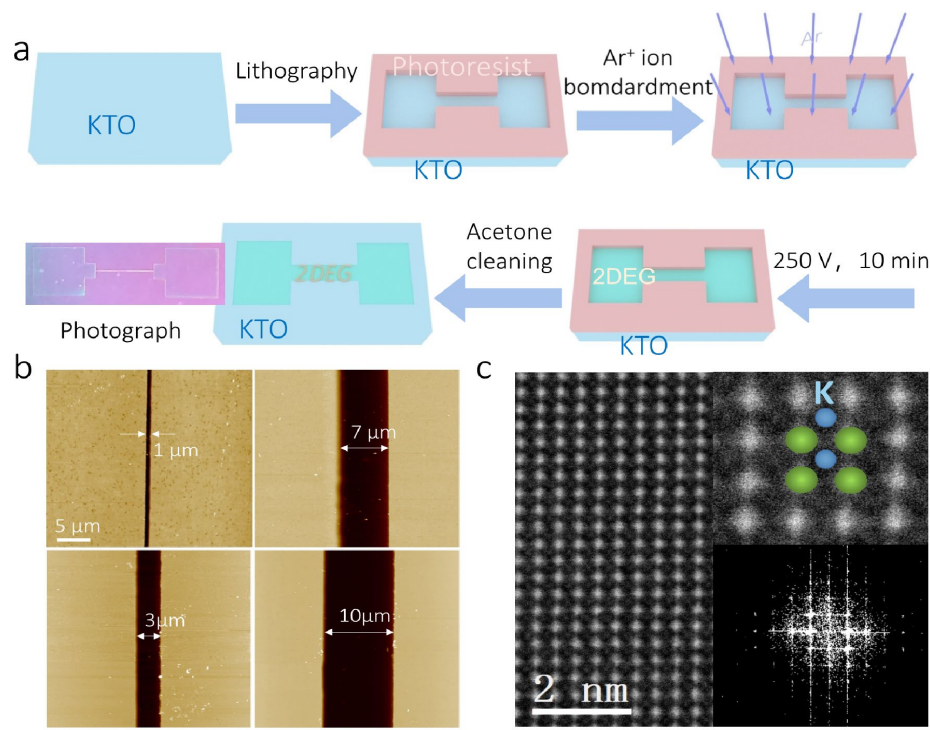

FIG. 1. (a) Schematic diagram of the preparation process of 2DEG channels using lithographic technique and Ar+-ion bombardment on a KTO substrate. (b) AFM images of the four 2DEG channels with the widths of 1 μm, 3 μm, 7 μm and 10 μm. (c) Cross-sectional STEM image showing the lattice of KTO along (001) direction. Inset (top): amplification of the lattice to exhibit the K (blue) and Ta (green) atoms. Inset (bottom): FFT pattern of KTO.

Here, we calculated the band structures of pristine KTO and oxygen-vacancy doped KTO (see Figs. S1 and S2 in the Supplemental Material). It is clear that the pristine KTO is an insulator with the band gap of 2.29 eV and 3.50 eV obtained by using the GGA-PBE+U (see Fig. S1c in the Supplemental Material) and HSE06 methods (see Fig. S1e in the Supplemental Material), respectively. The projected density of states (PDOS) plotted in Fig. S1d in the Supplemental Material shows that the valence bands and conduction bands of the pristine KTO are mainly contributed by O 2p states and Ta 5d states, respectively. In comparison, KTO becomes a metal with oxygen vacancies doped (see Fig. S2c in the Supplemental Material). One oxygen vacancy in 3×3×3 supercell of KTO will provide two electrons which makes the Fermi level cross over the conduction bands. The two electrons provided by the oxygen vacancy will fill Ta 5d states, as shown in the PDOS in Fig. S2d in the Supplemental Material. Therefore, oxygen vacancies on the surface of KTO can form a conductive nanoscale layer identified as a 2DEG. A KTO will undergo the phase transition from insulator to metal by introducing oxygen vacancies. The case of two oxygen vacancies in 3×3×3 supercell of KTO is also considered in Fig. S3 in the Supplemental Material, showing a similar metallic phase.

As a metal, a 2DEG always shows a decrease in resistance as the temperature drops[28]. Fig. 2a exhibits the temperature dependence of resistance of the 2DEG channels with different width. The 2DEG channel of 500 μm width shows a typical metallic conduction. With the channel width decreased to 10 μm, the resistance decreases first and then increases with the lowering of temperature, displaying a metal-insulator transition (MIT) at the transition temperature ($T_{tr}$) of 26 K. As the channel width decreases further, the insulator phase is dramatically enhanced and $T_{tr}$ shifts to a higher temperature. The 2DEG channel of 1 μm width can be regarded as an insulator with the resistance more than $10^{11}$ Ω below 10 K. It is intriguing that the MIT can be induced in KTO-based 2DEG through micron-scale size effect. This anomalous phenomenon can be attributed to the EPS. There are the separation and coexistence of metallic and insulating phases in the 2DEG region. Both of them exist as micron-sized islands or channels. For the wide enough 2DEG channel, metallic domains can always be connected together no matter how low the temperature is, thus giving rise to a metallic conduction. When the conductive channel is reduced into micron level, the metallic domains cannot communicate with each other due to the block of insulating domain. Therefore, the observed micron-scale size effect is the evidence of the EPS.

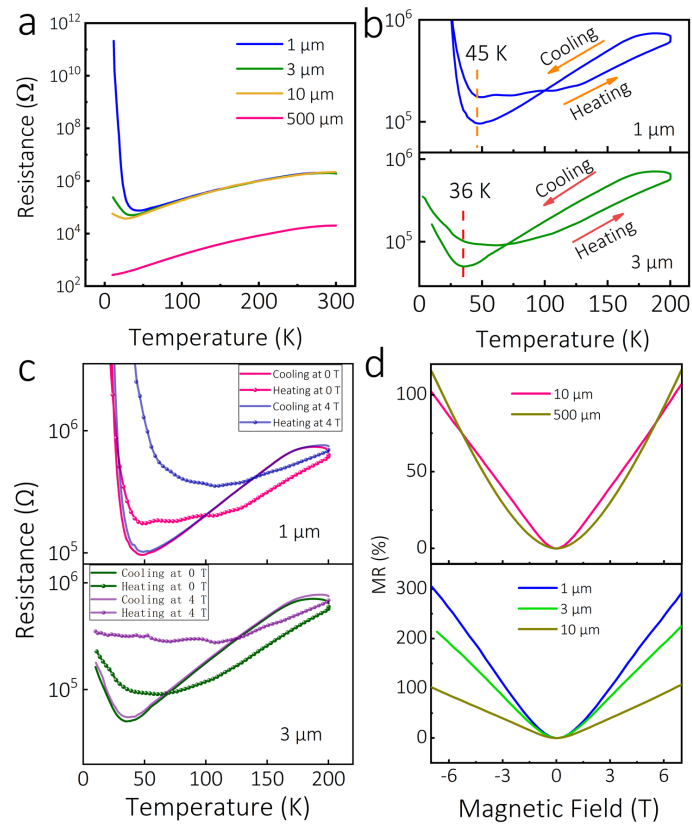

FIG. 2. (a) Resistance as a function of temperature for the 2DEG channels of 1, 3, 10 and 500 μm widths. (b) Resistance-temperature loops for the comparison between 1-μm and 3-μm 2DEG channels with temperature return line. (c) Resistance-temperature loops with and without a magnetic field of 4 T, for the 1-μm (top) and 3-μm (bottom) 2DEG channels. (d) MR as a function of magnetic field for the comparison between 10-μm and 500-μm 2DEG channels (top) and linear dependence of MR on the field for 1-μm, 3-μm and 10-μm 2DEG channels (bottom) at 25 K.

In Fig. 2b, we compare the resistance-temperature (R-T) relations upon heating and cooling in 1-μm and 3-μm 2DEG channels. It is observed that the heating curve does not overlap with the cooling one. Both of them show the same $T_{tr}$ despite the different resistance. Actually, the non-overlap of heating and cooling curves reflects the competition and relaxation characteristics during the MIT, implying the separation of metallic and insulating phases. It is possible that the domain transition between them induces the hysteretic R-T curves. Upon cooling, the insulating domains are continuously enlarged and change into the metallic ones around $T_{tr}$. An opposite physical process will occur upon heating. The transition from metallic to insulating domains may lag behind the temperature change, thus resulting in the R-T hysteresis behavior. Those domains tend to remain in their original phase state. This causes a smaller resistance upon cooling than that upon heating near $T_{tr}$. Based on the above understanding, $T_{tr}$ is just the temperature where the insulating domains start to separate the metallic domains and block the conductive channel. So the channel width actually corresponds to the maximum size of island-like insulating domain at the temperature of $T_{tr}$.

In addition, we consider the effect of magnetic field on the MIT. Fig. 2c shows the R-T loops of 1-μm and 3-μm 2DEG channels upon heating and cooling at a magnetic field of 4 T. There is a significant separation of heating curve from the cooling one. It is found that the magnetic field can make $T_{tr}$ shift to a higher temperature upon heating, differing from the case without field. Compared with the case of the cooling process, $T_{tr}$ is controllable by a magnetic field only during the heating process. It seems that the field prevents the transition from insulating to metallic domains, thereby causing higher $T_{tr}$. But it cannot affect the transition from metallic to insulating domains. In Fig. 3b, we compared the field dependences of resistance of 10-μm and 500-μm channels at 25 K. Here, MR is defined by MR=$(R_H-R_0)/R_0 \times 100\%$. It is obvious that the MR of 500-μm 2DEG channel depends on $H^2$, while the MR of 10-μm channel is linear with H. Such linear dependences of MR are also observed in 1-μm and 3-μm 2DEG channels. Giant magnetoresistance is observed in all of the channels. It is noted that the MR maximum increases significantly from 100% to 300% as the width decreases. Therefore, the EPS does not only induce the MIT, but also causes the unusual linear magnetoresistance behaviors.

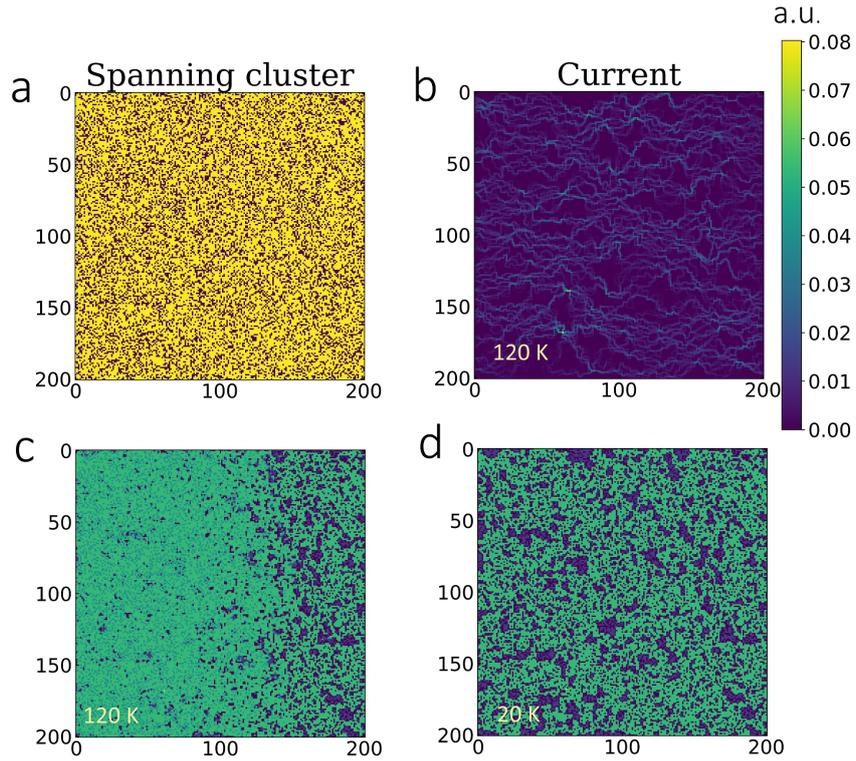

FIG. 3. (a) Randomly spatial distribution of oxygen vacancies in the $200 \times 200$ KTO lattice. "Yellow" and "black" sites correspond to the oxygen-vacancy occupied and unoccupied sites, respectively. (b) Simulated distribution of current paths at 120 K and at 1 V. BB (green) and DE (purple) mapping (c) at 120 K and (d) 20 K, respectively, showing the configuration of metallic and insulating domains.

To understand the anomalous MIT behavior in KTO-based 2DEG, we start from the first-principles calculations on the pristine and the oxygen-vacancy KTO. Based on density functional theory, the pristine KTO is an insulator with the band gap of 2.29 eV~3.50 eV (see Figs. S1 c and e in the Supplemental Material). For oxygen-vacancy KTO ($KTaO_{2.963}$ or $KTaO_{2.926}$), there exist finite density of states in the vicinity of the Fermi surface (see Fig. S2d and Figs. S3 d, h and i in the Supplemental Material). Such a band structure indicates the formation of metallic spanning-clusters on the surface of oxygen-vacancy KTO. Here, we introduce a simple percolation-type theoretical model to explain the observed MIT and EPS. The original oxygen ions in the pristine KTO form square lattice in the material, therefore we take a simple non-interacting electron square-lattice-crystal-structure model. In our simulations, the oxygen vacancies are stochastically spatially-distributed on the KTO surface, similar to the experimental case. We choose a series of different-size lattice, *e.g.* $L_x \times L_y = N_x a \times N_y a$ sites to simulate the experiments, where $a$ is the lattice constant, $N_x a$ and $N_y a$ represents the length of lattice in the longitudinal and vertical direction, respectively. Then, we take a certain probability of the sites to be occupied by oxygen vacancies, corresponding to the site-percolation. Fig. 3a exhibits a random distribution of the occupied sites (yellow) and unoccupied sites (black), representing the oxygen-vacancy and oxygen-ion regimes, respectively. Due to the metallic state of large-size 2DEG on KTO, a large probability (equal to 0.7) of oxygen-vacancies is used for our simulations. Based on the percolation theory (see Fig. 3b), a current distribution is simulated at 1 V and 120 K as illuminated in Fig. 3b. A large number of highly conductive paths are formed and corresponds to the metallic states in experiments. With a certain probability of the occupied-sites in the site-percolation, a spanning cluster is found in the system in Figs. 3 c and d. Backbone (BB) is defined as a set of all the sites that have at least two paths leading into them (one path from each side of the cluster). The remaining sites, which have only one path leading into them, are called as dangling ends (DE). We can get the BB as all the possible paths, including the necessary paths in the spanning cluster. A collection of BB paths corresponds to the metallic domains (green), while DE regions corresponds to the insulating domains (blue). In Fig. 3c, there are the coexistence of metallic and insulating domains (BB and DE), indicating an EPS. At 120 K, the metallic domains are dominant and the insulating domains are separated by them. With the temperature dropping into 20 K, the insulating domains grow larger, while the metallic domains are still connected with each other. In this case, despite the EPS, the sample maintain a metallic conduction in the whole temperature range, corresponding to the 500-μm 2DEG sample.

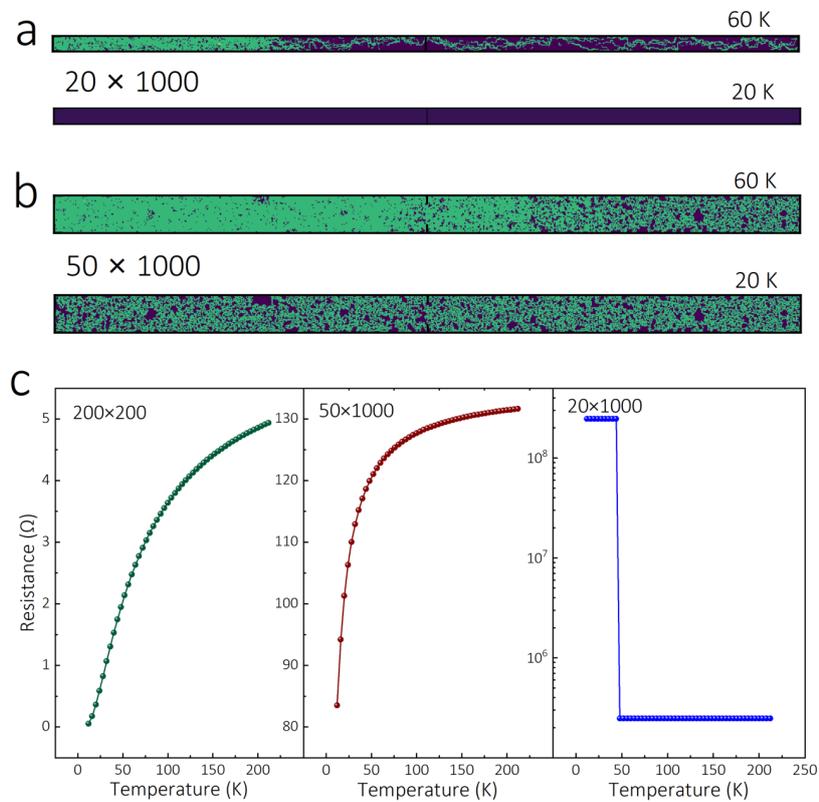

FIG. 4. BB (green) and DE (purple) mapping at 120 K and 20 K, for the (a) 20×1000 and (b) 50×1000 KTO lattices. (c) Simulated R-T curves for 200×200, 50×1000 and 20×1000 lattices, where each point corresponds to 50 different initial samples with arbitrary configurations.

In our experiments, the MIT shows a strong dependence on the channel width. Such a size-induced MIT can also be explained by our model. In Figs. 4 a and b, we simulate the distributions of BB and DE regions for 20×1000 and 50×1000 lattices. The lattices of $L_y \ll L_x$ are used to correspond to the narrow 2DEG channels in experiments. For the 20×1000 lattice, the BB paths are dense to form the conductive connection between two ends at 60 K, giving rise to a metallic conduction. The connection is cut off with the temperature dropping to 20 K, demonstrating the arising of insulating phase. In contrast, the 50×1000 lattice is full of BB paths even with the temperature decreasing to 20 K (see Fig. 4b), indicating a metallic conduction in the whole temperature range. Fig. 4c shows the R-T curves calculated through our model with different $L_y$. It is noted that the MIT emerges with width-length ration decreasing to 1/50. Our theoretical simulations show good agreement with our experimental results. It is clear that we have provided a useful platform to simulate the MIT phenomenon in a 2DEG system, especially for the disordered system originating from the oxygen-vacancy.

In summary, we fabricated the 2DEG channels with various widths through an $Ar^+$-ion bombardment on the surface of KTO. Such KTO-based 2DEGs show a size-dependent MIT effect, originating from the EPS. The observation of hysteretic R-T curves indicates the competition between metallic and insulating domains. Magnetic field has an effect on the competition process, causing different $T_{tr}$ between heating and cooling R-T curves. To explain the anomalous MIT, we propose a theoretical model based on the percolation theory. Our simulations support the separation of metallic and insulating phases at low temperature. With the width-length ratio decreasing, the insulating phase becomes dominant and the MIT emerges, showing good agreement with our experimental results.

Y.C.J. are grateful for the financial support from the National Natural Science Foundation of China (Grant No. 12274316) and Applied Fundamental Research Foundation of Nantong City, China (No. JC2021100). Y.Y. Z. acknowledges the support from the NSFC Grant No. 11804163 & Grant No. 12074276, and the Natural Science Foundation of the JiangSu Higher Education Institutions of China (Grant No. 18KJB140006). Y.Y.Z acknowledges the High-Performance Computing Center of Collaborative Innovation Center of Advanced Microstructures in Nanjing University, China.

**References**

1  Kagan, M. Y., Kugel, K. I. & Rakhmanov, A. L. Electronic phase separation: Recent progress in the old problem. *Physics Reports* **916**, 1-105 (2021).
2  Dagotto, E., Hotta, T. & Moreo, A. Colossal magnetoresistant materials: the key role of phase separation. *Physics Reports* **344**, 1-153 (2001).
3  Ward, T. Z. *et al.* Reemergent Metal-Insulator Transitions in Manganites Exposed with Spatial Confinement. *Phys. Rev. Lett.* **100**, 247204 (2008).
4  Zhu, Y. *et al.* Chemical ordering suppresses large-scale electronic phase separation in doped manganites. *Nat. Commun.* **7**, 11260 (2016).
5  Mohottala, H. E. *et al.* Phase separation in superoxygenated La2-xSrxCuO4+y. *Nat. Mater.* **5**, 377-382 (2006).
6  Kalcheim, Y. *et al.* Robust Coupling between Structural and Electronic Transitions in a Mott Material. *Phys. Rev. Lett.* **122**, 057601 (2019).
7  Miao, T. *et al.* Direct experimental evidence of physical origin of electronic phase separation in manganites. *Proc. Natl. Acad. Sci. U. S. A.* **117**, 7090-7094 (2020).
8  Zhai, H.-Y. *et al.* Giant Discrete Steps in Metal-Insulator Transition in Perovskite Manganite Wires. *Phys. Rev. Lett.* **97**, 167201 (2006).
9  McLeod, A. S. *et al.* Nano-imaging of strain-tuned stripe textures in a Mott crystal. *npj Quantum Materials* **6**, 46 (2021).
10  Dubi, Y., Meir, Y. & Avishai, Y. Nature of the superconductor–insulator transition in disordered superconductors. *Nature* **449**, 876-880 (2007).

11   Burgy, J., Mayr, M., Martin-Mayor, V., Moreo, A. & Dagotto, E. Colossal Effects in Transition Metal Oxides Caused by Intrinsic Inhomogeneities. *Phys. Rev. Lett.* **87**, 277202 (2001).

12   Strocov, V. N. *et al.* Electronic phase separation at LaAlO$_3$/SrTiO$_3$ interfaces tunable by oxygen deficiency. *Physical Review Materials* **3**, 106001 (2019).

13   Pavlenko, N., Kopp, T. & Mannhart, J. Emerging magnetism and electronic phase separation at titanate interfaces. *Phys. Rev. B* **88**, 201104 (2013).

14   Bert, J. A. *et al.* Direct imaging of the coexistence of ferromagnetism and superconductivity at the LaAlO3/SrTiO3 interface. *Nature Physics* **7**, 767-771 (2011).

15   Liu, C. *et al.* Two-dimensional superconductivity and anisotropic transport at KTaO3 (111) interfaces. *Science* **371**, 716-721 (2021).

16   Chen, Z. *et al.* Two-Dimensional Superconductivity at the LaAlO$_3$/KTaO$_3$ (110) Heterointerface. *Phys. Rev. Lett.* **126**, 026802 (2021).

17   Zou, K. *et al.* LaTiO3/KTaO3 interfaces: A new two-dimensional electron gas system. *APL Materials* **3**, 036104 (2015).

18   Zhang, H. *et al.* High-Mobility Spin-Polarized Two-Dimensional Electron Gases at EuO/KTaO$_3$ Interfaces. *Phys. Rev. Lett.* **121**, 116803 (2018).

19   Huang, G. *et al.* Time-dependent resistance of quasi-two-dimensional electron gas on KTaO3. *Appl. Phys. Lett.* **117**, 171603 (2020).

20   Shanavas, K. V. & Satpathy, S. Electric Field Tuning of the Rashba Effect in the Polar Perovskite Structures. *Phys. Rev. Lett.* **112**, 086802 (2014).

21   Kravchenko, S. V., Kravchenko, G. V., Furneaux, J. E., Pudalov, V. M. & D'Iorio, M. Possible metal-insulator transition at B=0 in two dimensions. *Phys. Rev. B* **50**, 8039-8042 (1994).

22   Kravchenko, S. V., Simonian, D., Sarachik, M. P., Mason, W. & Furneaux, J. E. Electric Field Scaling at a B=0 Metal-Insulator Transition in Two Dimensions. *Phys. Rev. Lett.* **77**, 4938-4941 (1996).

23   Popović, D., Fowler, A. B. & Washburn, S. Metal-Insulator Transition in Two Dimensions: Effects of Disorder and Magnetic Field. *Phys. Rev. Lett.* **79**, 1543-1546 (1997).

24   Lam, J., D'Iorio, M., Brown, D. & Lafontaine, H. Scaling and the metal-insulator transition in Si/SiGe quantum wells. *Phys. Rev. B* **56**, R12741-R12743 (1997).

25   Das Sarma, S. *et al.* Two-Dimensional Metal-Insulator Transition as a Percolation Transition in a High-Mobility Electron System. *Phys. Rev. Lett.* **94**, 136401 (2005).

26   Meir, Y. Percolation-Type Description of the Metal-Insulator Transition in Two Dimensions. *Phys. Rev. Lett.* **83**, 3506-3509 (1999).

27   Reagor, D. W. & Butko, V. Y. Highly conductive nanolayers on strontium titanate produced by preferential ion-beam etching. *Nat. Mater.* **4**, 593-596 (2005).

28   Ohtomo, A. & Hwang, H. Y. A high-mobility electron gas at the LaAlO3/SrTiO3 heterointerface. *Nature* **427**, 423-426 (2004).

# Supplemental Material

# Metal-insulator phase separation in $KTaO_3$-based two-dimensional electron gas

*Jinlei Zhang[1,#], Jiayong Zhang[1,#], Dapeng Cui[2,#], Li Ye[1,#], Shuainan Gong[1], Zhichao Wang[1],*

*Zhenping Wu[3], Chunlan Ma[1], Ju Gao[1,4], Yuanyuan Zhao[5,]\* and Yucheng Jiang[1,]\**

[1]Jiangsu Key Laboratory of Micro and Nano Heat Fluid Flow Technology and Energy Application, School of Physical Science and Technology, Suzhou University of Science and Technology, Suzhou, 215009, China

[2]Department of Physics, Faculty of Science, National University of Singapore, Singapore 117551, Singapore

[3]State Key Laboratory of Information Photonics and Optical Communications & School of Science, Beijing University of Posts and Telecommunications, Beijing 100876, China

[4]School for Optoelectronic Engineering, Zaozhuang University, Shandong 277160, China

[5] School of Physics and Optoelectronic Engineering, Nanjing University of Information Science and Technology, Nanjing 210044, China.

#These authors contributed equally to this work.

*Email: yyzhao@nuist.edu.cn and jyc@usts.edu.cn

**Methods and figures**

**Methods**

**Fabrication of 2DEG channels.** A (001) KTO substrate was covered by photoresist, and then the suitable channel patterns were selected for lithography (see Fig. 1a). The resulting samples were placed into a chamber and bombarded by Ar+ beam for 10 min under the voltage of 250 V. Ar gas pressure and flow rate are $3\times10^{-4}$ mbar and 6 sccm, respectively. After the etching, the remaining photoresist on the sample surface was cleaned with acetone and deionized water. The KTO-based 2DEG is unstable in the air, which may be destroyed by oxygen. The sample is coated by polymethyl methacrylate to prolong the lifetime of conductive channels.

**Characterization and measurements.** An AFM technique was used to probe the surface morphology and widths of 2DEG channels (MultiMode 8, Bruker). Cs-corrected STEM was used to investigate the lattice structure of (001) KTO. Two wires were bonded to the terminals of 2DEG channels by using Manual Wire Bonders ibood5000. Keithley 6517b Source Meter was used to apply the voltages and measure the resistance. All electrical measurements were performed in a physical property measurement system (Quantum Design) with a helium atmosphere.

**First-principles calculations.** The first-principles calculations of the KTO are carried out by using the projected augmented wave (PAW) formalism based on density functional theory (DFT) [1], as implemented in the Vienna ab-initio simulation package (VASP) [2]. The Perdew-Burke-Ernzerhof generalized-gradient approximation (GGA-PBE) is employed to describe the exchange and correlation functional [3]. All of the structural optimizations for the KTO unit cell and supercell are calculated by using the SCAN functional [4]. The lattice constant of KTO unit cell obtained by the SCAN functional is 3.993 Å, which is in good agreement with experiments (3.989 Å [5]). The plane-wave cutoff energy is set to be 500 eV. The convergence criterion for the total energy is set to be $10^{-6}$ eV. All atoms in the unit cell are allowed to relax until the Hellmann-Feynman force on each atom is smaller than 0.02 eV/Å. The Gamma central Monkhorst-Pack k-point grids of $12\times12\times1$ and $4\times4\times1$ are employed to sample the first Brillouin zone of the unit cell and $3\times3\times3$ supercell of KTO, respectively. To investigate the band structures and density of states of the pristine KTO and KTO with oxygen vacancy, the GGA+U method [6] is adopted to describe the correlation effects of the Ta 5d electrons, and the values of the Coulomb interaction U and exchange interaction J are set to be 2.0 and 1.0 eV, respectively. The band structure of the pristine KTO unit cell is also calculated by using the HSE06 functional [7]. The climbing-image nudge elastic band (CI-NEB) method [8] is used to calculate the oxygen vacancy migration pathway and energy barrier in the $3\times3\times3$ supercell of KTO.

**Percolation model.** Several 2D MIT have been observed in various systems since the 1990s, for example, the Si-metal-oxide-semiconductor field-effect transistor (MOSFET) inversion layers [9-12]. Many theories have been developed to explain the phenomena considering various effects in real systems [13]. Among the explanations, a percolation-type description of the 2D-MIT [14-21] has been suggested. Such a percolation transition for 2D systems was discussed by A. L. Efros [14], J. A. Nixon and J. H. Davies [15]. Here we introduce a simple percolation-type model [22-25] to explain the phenomenological 2D MIT in our experiments. The percolation transition for 2D MIT is highly physically motivated, and our experiments could be explained qualitatively in a simple way. We neglect other physical effects on 2D MIT, such as the disorder effect, electron-electron interaction and scattering by impurities [13].

In the following, we describe our model and simulations in detail. The original oxygen ions in the pristine KTO form a square lattice in the material, so we perform a simple non-interacting electron square-lattice-crystal-structure model. We can choose a series of different-size lattices in our simulation: $L_x \times L_y = N_x a \times N_y a$ sites, with $a$ the lattice constant. We can simply take it as a unit $a = 1$. $N_x a$ and $N_y a$ represents the length of lattice in the longitudinal and vertical direction, respectively. The oxygen-vacancies in our simulation correspond to the sites randomly occupied in our lattice, similar to the site-percolation [22-25]. Our experiments indicate that the large-size quasi-2DEG KTO surface exhibits metallic properties. Thus, we choose a large probability (Q_r) of oxygen-vacancies, *e.g.,* we choose $O_r = 0.70$ in the simulation.

The oxygen-vacancies 2D KTO system is a disordered material, and especially, it can be described to be the site-percolation lattice. To understand the conductivity here, we can proceed the site-percolation problem to a bond-percolation network. Afterwards, the Random Resistor Networks could be applied to study the conductive properties of such a bond-percolation network [25-28]. A voltage $V$ is applied across the disordered material. We choose this voltage along the longitudinal direction in our simulation, and take $V = 1$ as the unit. Then we choose a bond conductance between two adjacent sites as $G_{<ij>}$. In the metallic regime and depletion regime, the conductivity is distinct. In the metallic regime, the carriers can move easily and smoothly, while in the depletion regime, the carriers are hard to move. Thus, we take $G_{<ij>-metallic}$ and $G_{<ij>-depletion}$ along the adjacent site $i$ and $j$ in the metallic and depletion regime, respectively. Once we choose the voltage in the longitudinal direction and the bond conductance $G_{<ij>}$, we can simulate to get current $I$ with Kirchhoff's equations. Based on the simulations, we can get the random 'hill-and-valley' voltage-potential landscape, and then the current along each adjacent bond and the total current could be obtained.

The conductance of a classical 2DEG originates from the Arrhenius formula [28]. We take $G_{<ij>} = g \exp\left(\frac{\Delta_{gap}}{2k_B T}\right)$, where $T$ is the temperature, $k_B \sim 8.62 \times 10^{-5}$ eV/K is the Boltzmann constant and $g$ is a factor. We can easily choose $g_{metallic} = 1$ as a unit in the metallic regime, while in the depletion regime $g_{depletion} = 10^{-6}$. Then $\Delta_{gap}$ is obtained from the band structure, at an order of 0.01 eV (we take $\Delta_{gap} = 0.02$ eV). This small value favors the percolation property, otherwise the system will exhibit metallic properties, without a bond-percolation.

Whether the carriers move along each bond between two adjacent sites depends on the kinetic energy and the potential. To mimic the process, we construct a relative probability of carrier transmission $\left(P_{NN} \sim \frac{1}{\exp\left(\frac{E-E_f}{k_B T}\right)+1}\right)$ along an adjacent bond (nearest-neighbor site), which is based on the Fermi-Dirac distribution, where $E$ represents the energy of the electron. It is straightforward to take $E$ from our first-principles calculations. In the metallic regime, $E_{metallic} \sim 0 - 0.05$ eV, moreover, $E_{metallic} = 0.05$ eV × a uniformly distributed random number in $[0,1]$; while in the depletion regime $E_{depletion} = -3$ eV; and $E_f$ represents the Fermi energy, taken to be 0; $T$ is the temperature, and $k_B \sim 8.62 \times 10^{-5}$ eV/K is Boltzmann constant. Furthermore, the carriers will be frozen gradually while the temperature cools down. Below a certain temperature $T_{frozen} = 10$ K, most of the carriers will be frozen. The whole procedure originates from thermal activation and tunneling. With all these pre-preparations, we choose $L_x \times L_y$ = 200×200, 20×1000 and 50×1000 lattices as our model, and then choose temperatures ranging from 200 to 20 K in our simulation. At the starting temperature $T_{starting} = 200\text{ K}$, we choose 200 different initial site-percolation configurations. Each configuration evolves with the temperature cooling down. At each temperate, we simulate 50 round random-judgment searches over the complete lattice. Finally, we get the statistical average of the resistance of the system at each temperature.

**References**

[1] P. E. Blöchl, Projector augmented-wave method, Phys. Rev. B **50**, 17953 (1994).

[2] J. P. Perdew, K. Burke, and M. Ernzerhof, Generalized Gradient Approximation Made Simple, Phys. Rev. Lett. **77**, 3865 (1996).

[3] J. P. Perdew, K. Burke, and M. Ernzerhof, Generalized Gradient Approximation Made Simple, Phys. Rev. Lett. **77**, 3865 (1996).

[4] J. W. Sun, A. Ruzsinszky, and J. P. Perdew, Strongly Constrained and Appropriately Normed Semilocal Density Functional, Phys. Rev. Lett. **115**, 036402 (2015).

[5] S. H. Wemple, Some Transport Properties of Oxygen-Deficient Single-Crystal Potassium Tantalate ($KTaO_3$), Phys. Rev. **137**, A1575 (1965).

[6] A. I. Liechtenstein, V. I. Anisimov, and J. Zaanen, Density-functional theory and strong interactions: Orbital ordering in Mott-Hubbard insulators, Phys. Rev. B **52**, R5467(R) (1995).

[7] J. Heyd, G. E. Scuseria, and M. Ernzerhof, Hybrid functionals based on a screened Coulomb potential, J. Chem. Phys. 118, 8207 (2003); 124, 219906 (2006).

[8] G. Henkelman, B. P. Uberuaga and H. Jonsson, A climbing image nudged elastic band method for finding saddle points and minimum energy paths, J. Chem. Phys., 113, 9901–9904 (2000).

[9] S. V. Kravchenko, W. E. Mason, G. E. Bowker, J. E. Furneaux, V. M. Pudalov, and M. D'Iorio, Phys. Rev. B. 51, 7038 (1995)

[10] V. M. Pudalov, G. Brunthaler, A. Prinz & G. Bauer, JETP Lett. 65, 932 (1997).

[11] D. Popović, A. B. Fowler, and S. Washburn, Phys. Rev. Lett. 79, 1543 (1997).

[12] J. Lam, M. D'Iorio, D. Brown, and H. Lafontaine, Phys. Rev. B 56, R12741(R) (1997).

[13] E. Abrahams, S. V. Kravchenko, and M. P. Sarachik, Rev. Mod. Phys. 73, 251 (2001).

[14] A. L. Efros, Solid State Commun. 65, 1281(1988), A. L. Efros, Solid State Commun. 70, 253(1989).

[15] J. A. Nixon and J. H. Davies, Phys. Rev. B 41, R7929 (1990).

[16] J. Shi and X. C. Xie, Phys. Rev. Lett. 88, 086401 (2002).

[17] S. He and X. C. Xie, Phys. Rev. Lett. 80, 3324 (1998).

[18] Y. Meir, Phys. Rev. Lett. 83, 3506 (1999).

[19] R. Leturcq, D. Lʼhôte, R. Tourbot, C. J. Mellor, and M. Henini, Phys. Rev. Lett. 90, 076402 (2003).

[20] M. M. Fogler, Phys. Rev. B 69, 121409 (2004).

[21] S. Das Sarma, M. P. Lilly, E. H. Hwang, L. N. Pfeiffer, K.W. West, and J. L. Reno, Phys. Rev. Lett. 94, 136401 (2005).

[22] D. Stauffer, A. Aharony, Introduction to Percolation Theory, Taylor & Francis, 2nd Edition, London, (2017).

[23] B. I. Shklovskii, A. L. Efros, Electronic Properties of Doped Semiconductors,

Springer Berlin, Heidelberg (1984).

[24] J. W. ESSAM, Rep. Prog. Phys. 43 833 (1980).

[25] Anders Malthe-Sørenssen, https://www.uio.no/studier/emner/matnat/fys/FYS4460/v20/notes/book.pdf

[26] B. J. Last and D. J. Thouless, Phys. Rev. Lett. 27, 1719 (1971).

[27] S. Kirkpatrick, Rev. Mod. Phys. 45, 574 (1973).

[28] J. Adler, Y. Meir, A. Aharony, A. B. Harris & Lior Klein, J Stat Phys 58, 511–538 (1990).

**Supplemental Figures**

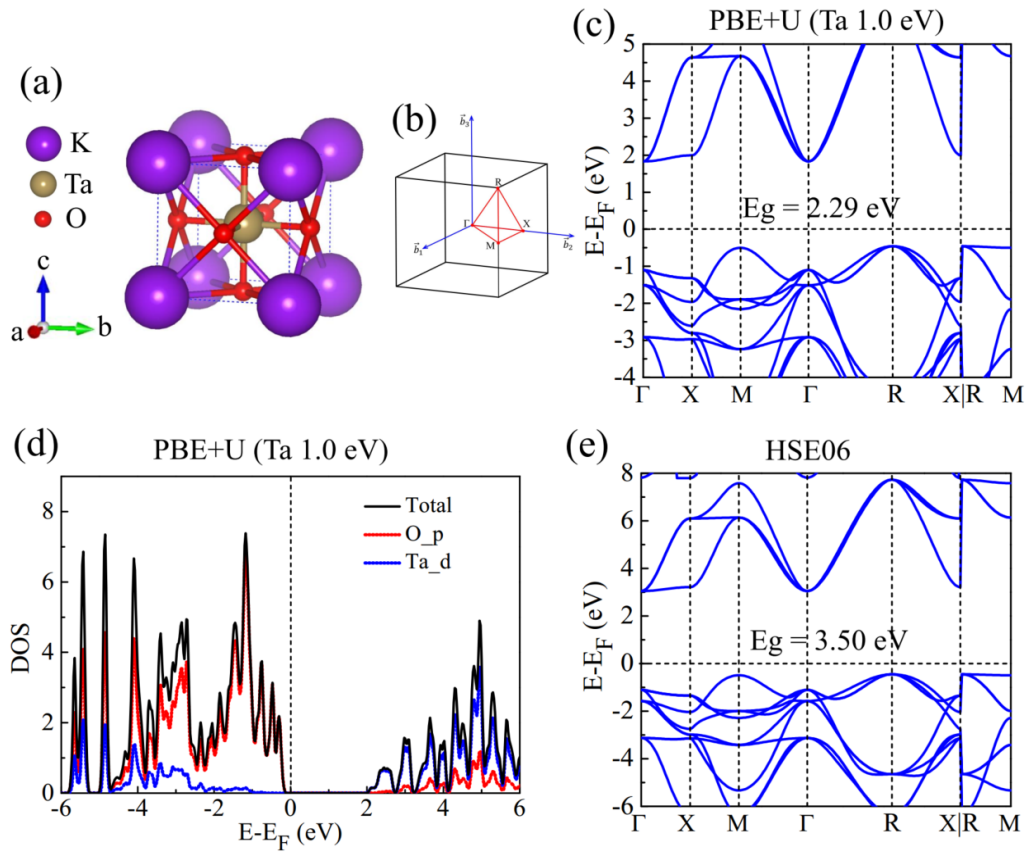

FIG. S1. (a) The atomic structure for the unit cell of the KTO. (b) The first Brillouin Zone and the special k points with high symmetries for the KTO structure. (c) The calculated band structure of KTO until cell by using the GGA-PBE+U method with U = 1.0 eV. (d) The projected density of states (PDOS) of the pristine KTO structure. (e) The calculated band structure of KTO until cell by using the HSE06 functional.

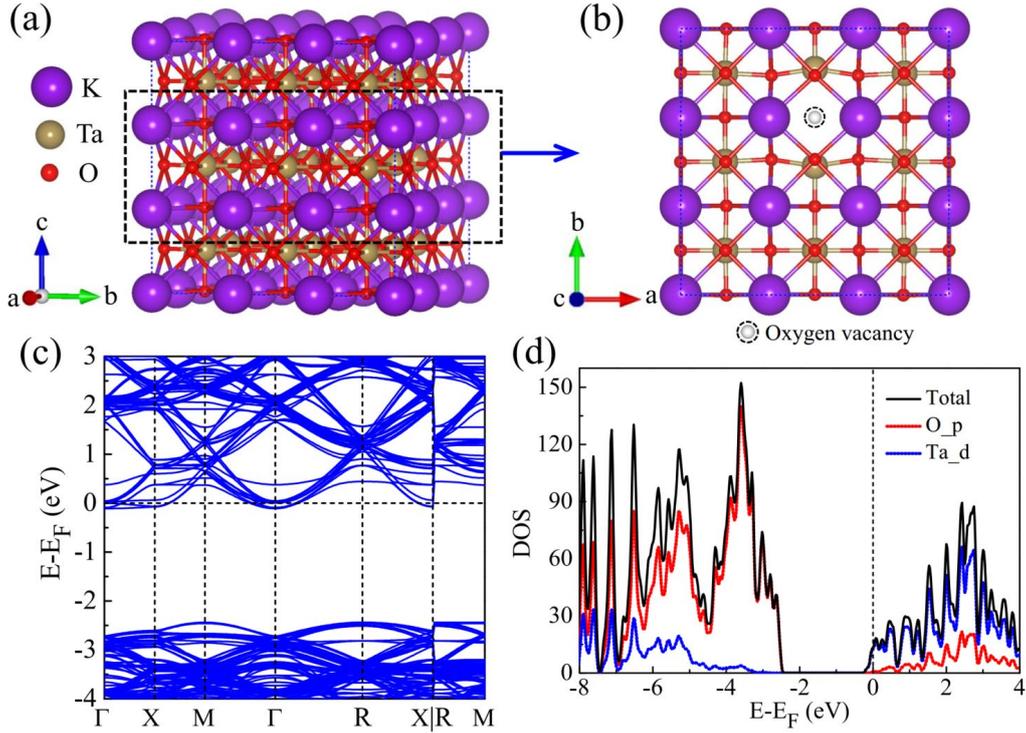

FIG. S2. (a) The atomic structure of the 3×3×3 supercell of KTO with one oxygen vacancy. (b) The magnified atomic structure in the rectangular dashed box in (a). The oxygen vacancy is denoted in (b). (c) The calculated band structure of the 3×3×3 supercell of KTO with one oxygen vacancy (KTaO$_{2.963}$) by using the GGA-PBE+U method with U = 1.0 eV. (d) The corresponding projected density of states for the 3×3×3 supercell of KTO with one oxygen vacancy (KTaO$_{2.963}$).

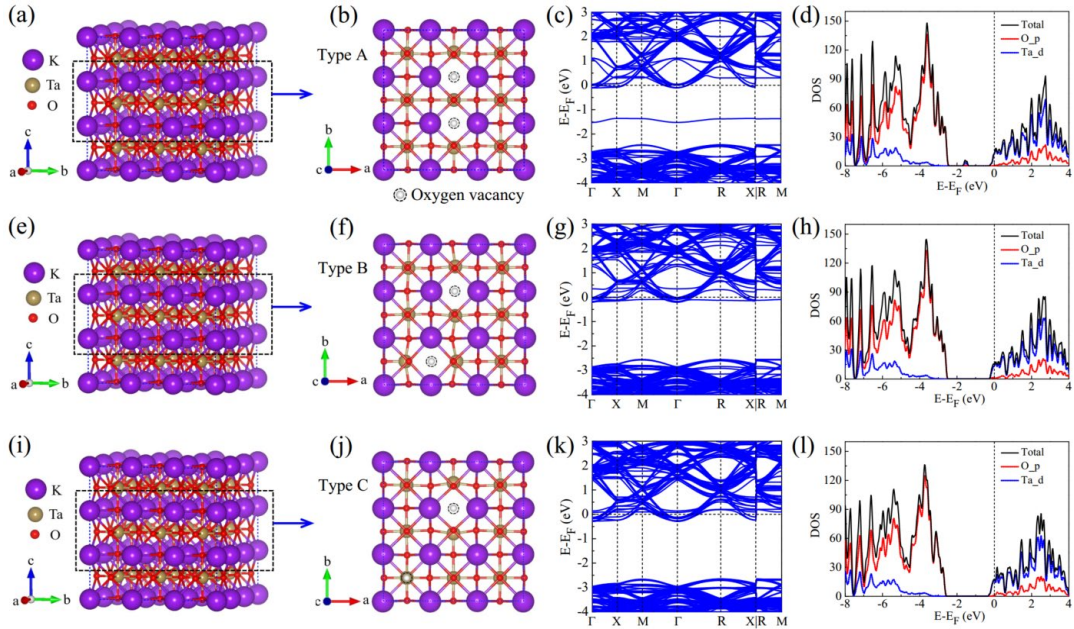

FIG. S3. (a) The atomic structure of the 3×3×3 supercell of KTO with two oxygen vacancies. (b)

The magnified atomic structure in the rectangular dashed box in (a), where the configuration for the two oxygen vacancies in (b) is defined as Type A. (c,d) The calculated band structure (c) and projected density of states (d) of the 3×3×3 supercell of KTO with two oxygen vacancies (KTaO$_{2.926}$) of Type A configuration by using the GGA-PBE+U method with U = 1.0 eV. (e)-(h) The same as (a)-(d), except that the calculated system is the 3×3×3 supercell of KTO with two oxygen vacancies (KTaO$_{2.926}$) of Type B configuration. (i)-(l) The same as (a)-(d), except that the calculated system is the 3×3×3 supercell of KTO with two oxygen vacancies (KTaO$_{2.926}$) of Type C configuration.